\documentclass[twocolumn]{article}
\usepackage{graphicx}
\usepackage{authblk}
\usepackage[left=2cm,right=2cm]{geometry}
\usepackage{amsmath, amscd, amssymb, amsthm}

\usepackage{caption}
\usepackage{subcaption}
\usepackage[title]{appendix}
\usepackage{multirow}

\title{Plasma Confinement State Classification in Fusion Power Plants: Profile Reflectometer and Ensemble Diagnostics}
\author[1]{Randall Clark} 
\author[2]{Vacslav Glukhov} 
\author[2]{Georgy Subbotin}
\author[2]{Maxim Nurgaliev} 
\author[2]{Aleksandr Kachkin} 
\author[3]{Lei Zeng} 
\author[1]{Dmitri M. Orlov}
\affil[1]{Center for Energy Research, University of California San Diego, La Jolla, CA, 92093}
\affil[2]{Next Step Fusion S.a.r.l., Luxembourg}
\affil[3]{University of California Los Angeles, Los Angeles, CA, 90095}

\date{December 2025}

\begin{document}


\twocolumn[
  \begin{@twocolumnfalse}
    \maketitle
    \begin{abstract}
     As Fusion Pilot Plants (FPPs) are increasingly viewed as within reach, many engineering challenges remain. Not many diagnostics are expected to be available in a reactor environment. Survivability, maintainability, and limited port space substantially restrict the number of FPP-relevant diagnostics. One remaining challenge is developing tools and devices to extract plasma state information necessary for controlling an FPP from a limited subset of diagnostics. This work is part of an overarching project to address this challenge. The specific diagnostic subset to be used in FPPs is still under debate.  We take the approach of developing machine-learning-based tools for different significant plasma state parameters, using already known FPP-viable diagnostics. Previously we developed a plasma confinement mode classifier utilizing the Electron Cyclotron Emission (ECE) diagnostic \cite{clark2025plasma}. Here, we expand on this by developing a Profile Reflectometer (PR) based classifier with 97\% test accuracy, and an ensemble model that combines the ECE and PR models into a single model, achieving 99\% test accuracy. 
    \end{abstract}
  \end{@twocolumnfalse}
]

\section{Introduction}
The growing focus on fusion power plant (FPP) design and its limitations has raised serious questions about the relevant diagnostics. Chief among them are the diagnostics' ability to operate in the harsh conditions of fusion power generation, and their information capacity for plasma control. The need for survivability, low maintenance and replacement costs, and limited port window size are significant constraints that limit the set of relevant diagnostics. Plasma control informed by a severely constrained set of diagnostics thus represents an FPP-specific design challenge. In this paper, we continue reporting the development of plasma state identification using FPP-relevant diagnostics, this time, using the profile reflectometer at DIII-D. 

Plasma confinement mode is one of the most important attributes of the state of a fusion plasma.  The high confinement mode (H-mode), as opposed to the low confinement mode (L-mode), is to be the primary mode of operation in FPPs. Next generation tokamaks like ITER, SPARC, and DEMO are all expected to run in H-mode \cite{rodriguez2022overview,mukhovatov2003overview,siccinio2022development}. The distinct characteristic of H-mode is the pedestal, which sharply raises density and temperature.  

Existing H-mode classifiers rely on a broad set of research-focused diagnostics that will not be available in a reactor environment. We have begun addressing this critical gap in our prior work, in which we used the Electron Cyclotron Emission (ECE) diagnostic to develop robust and efficient ML methods to identify the plasma confinement mode \cite{clark2025plasma}. FPP design efforts benefit from expanding the set of relevant diagnostics and plasma state models. The focus of the paper is to investigate whether the profile reflectometer (PR) diagnostic at DIII-D can also accurately and reliably identify the plasma confinement mode. We also develop an ensemble model combining both ECE and PR diagnostics to achieve even greater accuracy and robustness.

\section{Diagnostics}
\subsection{The Profile Reflectometer Diagnostic}
Frequency-modulated continuous wave (FM-CW) reflectometry has been widely employed for the measurement of electron density profile ($n_e$) in fusion studies. It is a short-range radar-like technique, measuring either the probe wave time delay or phase shift from the plasma cutoff layers. The phase shift is a line integrated function of refractive index, $\mu(x,f)$, represented as:

\begin{equation}
\Phi(f)=\frac{4\pi f}{c} \int_{x_c(f)}^{x_0}\mu(x,f)  dx-\pi/2
\end{equation}

Where $f$ is reflectometry frequency, $x_0$ is the plasma starting position and $x_c(f)$ is the plasma cutoff layer. By using the digital complex demodulation (CDM) technique, the phases are extracted from the reflectometry signal.  Because $\mu$ is related to $n_e$, the density profile can be inverted and reconstructed numerically from $\Phi(f)$ \cite{kim1997development}.  

In DIII-D, profile reflectometry system routinely operates with dual-polarization for both Q-band (34–50 GHz) and V-band (49–75 GHz) frequency bands \cite{zeng2006fast}. Thereby,  the $n_e$ range of 0 to 7x$10^{19}$ $m^{-3}$ can be measured with high temporal resolution (25 $\mu$s). It has been employed to study fast physics events in plasmas, such as L-H mode transition, QH-mode, MHD, and Internal Transport Barriers. Although this diagnostic has many advantages, there are some limitations. First the operational RF frequency range determines ne measurement coverage. Second,  the magnetic field  in the machine should be higher than a certain value (typical 1.6 T in DIII-D) in order to detect the first right-hand cutoff location in plasma. Third, the cutoff layer should be monotonic, not a hollow profile. 

It is believed that it can be employed in future fusion reactor and FPP due to it being low cost and compact option. It only requires limited spatial access with no requirements for neutral beam injection and is capable of high time and spacial resolution. The hazardous environment of a reactor won't impact it due to its sensitive instruments being located outside the tokamak walls.


\subsection{The Electron Cyclotron Emission Diagnostic}
The Electron Cyclotron Emission (ECE) diagnostic measures the electron temperature profile and has previously been used in the ECE-based H-mode classifier \cite{clark2025plasma}. In a magnetized plasma, electrons will gyrate at the cyclotron harmonics. Plasma is optically thick at the corresponding frequencies and emits approximately as a black body. The measured intensity of light is therefore proportional to the temperature of the electron. In the ECE diagnostic, as in other microwave diagnostics, radiation is collected at the plasma boundary and transported via waveguides to a detection instrument protected from the reactor's harsh environment \cite{austin2003,hartfuss2013mwdiag}. 
\section{Data Analysis}
\subsection{Hand Label Generation Process}
The L- and H-mode-labeled data used to train and test the PR model are the same as those we previously used for the development of the ECE-based confinement mode identification model. They consist of 300 shots collected across 2024 and 2025, spanning many different experimental days to explore a wide range of plasma parameters. We refer the reader to our published work on the ECE-based classifier, which details the properties of the labeled data  \cite{clark2025plasma}. 

\subsection{Peculiarities of Profile Reflectometer data}

The Profile Reflectometer data are not readily available for all shots. Achieving acceptable data quality involves a post-measurement workflow that requires manual intervention. Due to this complication, the original labeled set is reduced from 300 to 260 shots for which PR data are available for training and testing. Additionally, density profiles are not always available for the entire shot period, and the start time of PR data varies, further reducing the available data for our model. Since every shot begins in L-mode, the consequence of the PR data not always starting at t=0 ms is that the set is biased towards the H-mode data. Time-slices of data were taken every 100 ms from the 260 available shots, with PR data comprising 8102 samples for training and testing; of these, 66\% were H-mode and 34\% were L-mode. 

Another challenge, in contrast to the ECE diagnostics, stems from the physics and limitations of PR density measurements: the observed profile does not always extend over the full range of the plasma from the plasma core to the plasma edge. As the PR performs frequency sweeps, its maximum injected frequency sets a hard constraint on the observable density. Once this upper limit is reached, the PR can no longer penetrate deeper into the plasma. For this reason, a non-negligible fraction of profile measurements stop short of the core $\rho \simeq 0$, sometimes measuring only the pedestal region and the outer plasma edge $\rho \simeq 1$. We address this challenge in the model construction section that follows. 

\begin{figure}[ht!]
\centering
\includegraphics[width=8.5cm,height=8.5cm,keepaspectratio]{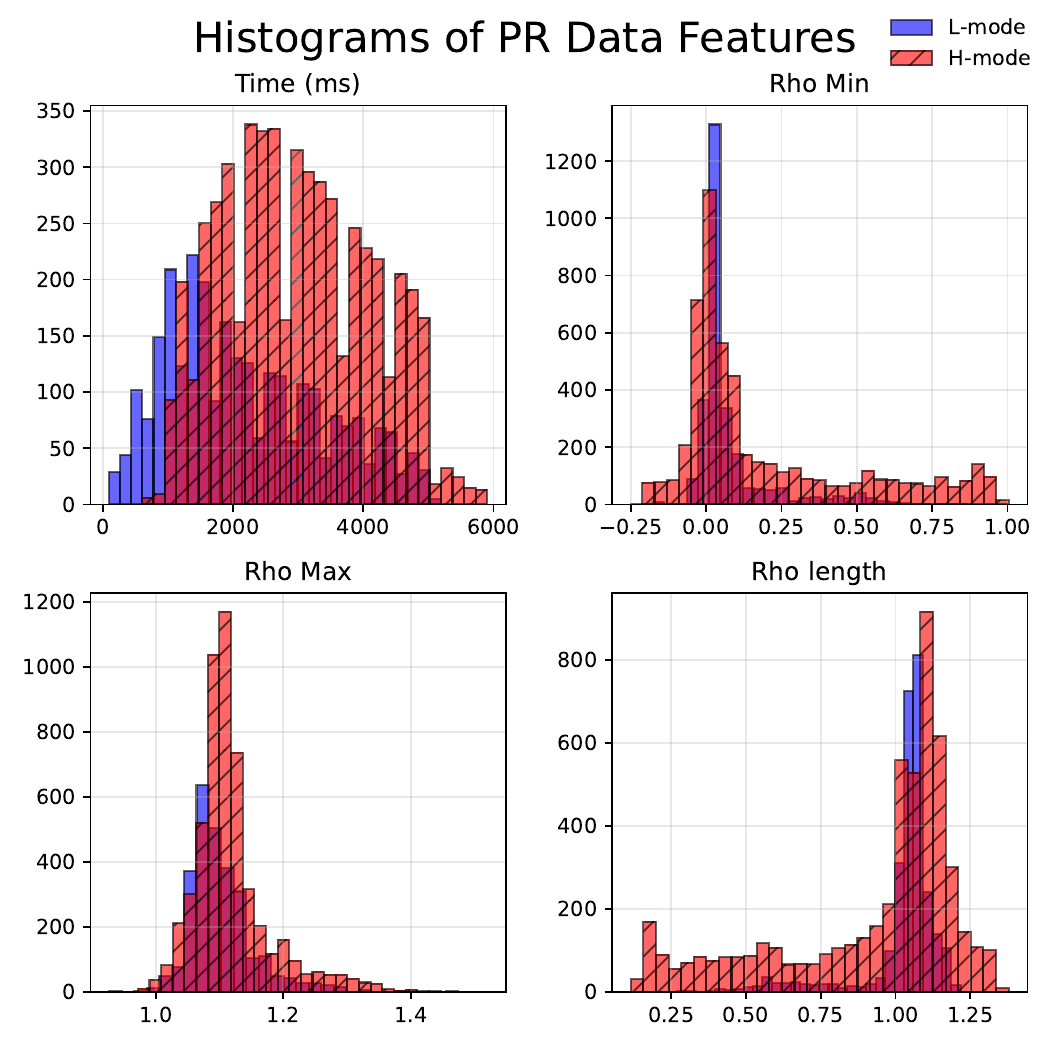}
\caption{Distribution of the PR data used in training and testing according to the time within the shot, and the $\rho$ range. }
\label{Histogram}
\end{figure}

In Figure \ref{Histogram}, we show the distribution of L- and H-mode labeled data in the data set. It clearly demonstrates that the H-mode labeled data almost never start at $t=0$ and that the L-mode data most of the time cover the whole interval of $\rho$. It is also noteworthy that the H-mode data points dominate large $\rho \simeq 1$. In the following section, we explain how these properties are incorporated into the model. 

\begin{figure}[!ht]
\centering
\includegraphics[width=8.5cm,height=8.5cm,keepaspectratio]{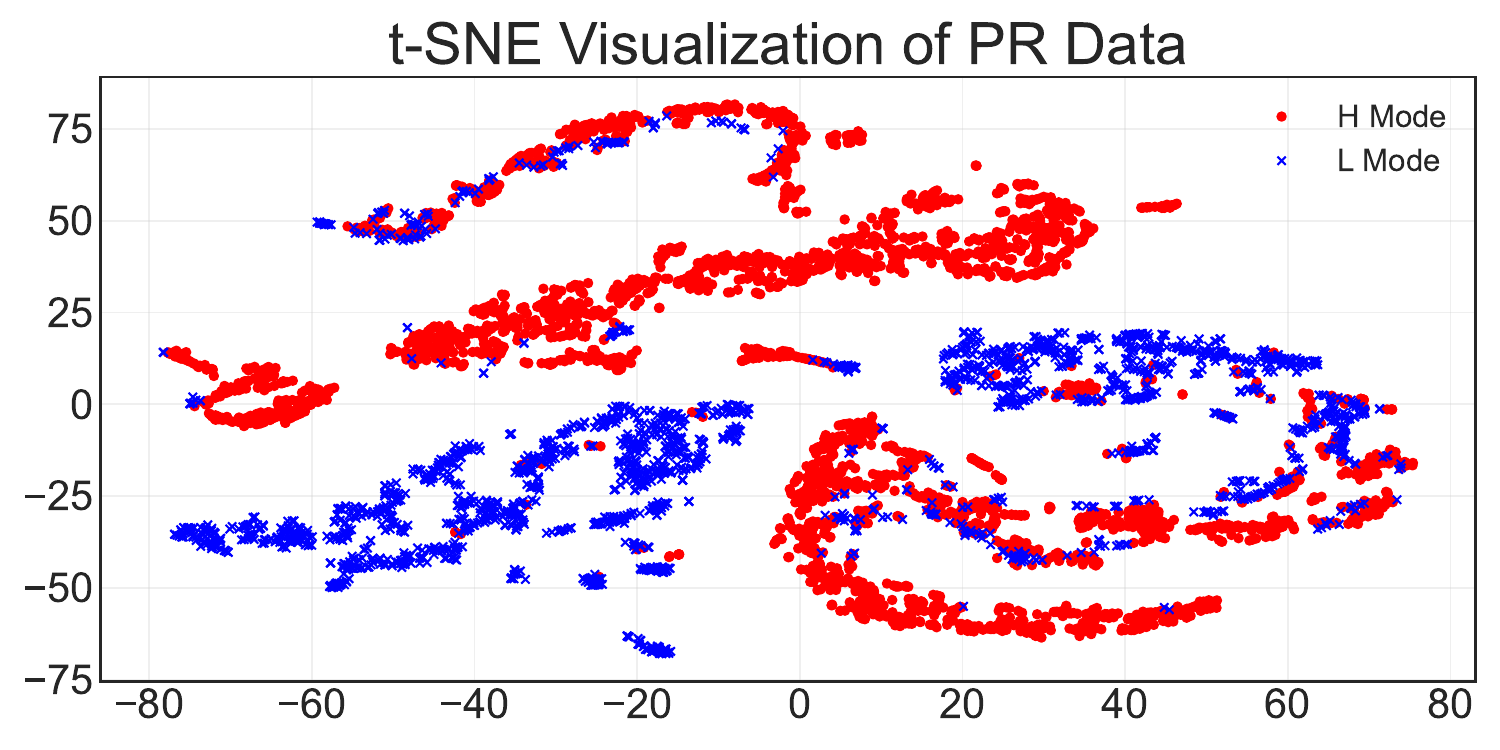}
\caption{A t-SNE heuristic visualization of PR labeled data. Apart from a few outliers at the boundary between L- and H-mode, the hand-labeled data points separate cleanly.}
\label{TSNE}
\end{figure}

Figure \ref{TSNE} shows the t-SNE \cite{JMLR:v9:vandermaaten08a} visualization of the L- and H-mode separation in the raw feature space. The t-SNE heuristic is a visualization and dimensionality reduction technique that preserves local clusters of similar behavior. The clear separation of the two models in this chart assures that the PR-based confinement mode classification is feasible. 

\section{The Profile Reflectometer Classifier}
\subsection{The Model}

A distinct characteristic of the PR diagnostic is the inability of the probing electromagnetic wave to penetrate the plasma core once its local density exceeds a threshold.  On the other hand, the physics of the high-confinement mode, characterized by the presence of a pedestal, indicates that the edge region is most important for classification -- here, we refer the reader to our previous work \cite{clark2025plasma}. For this reason, a robust feature extraction method needs to handle the full depth of the plasma when the corresponding data are available, but also gracefully handle situations when the core data are not available. 

After experimentation, we concluded that most of the profiles can be well fit by 3rd-order polynomial splines with 10 knots placed throughout the density data \cite{de1978practical}. The feature extraction method smooths out occasional irregularities in PR data and interpolates the density at arbitrary $\rho$ where data are available. 

\begin{figure}[ht!]
\centering
\includegraphics[width=8.5cm,height=8.5cm,keepaspectratio]{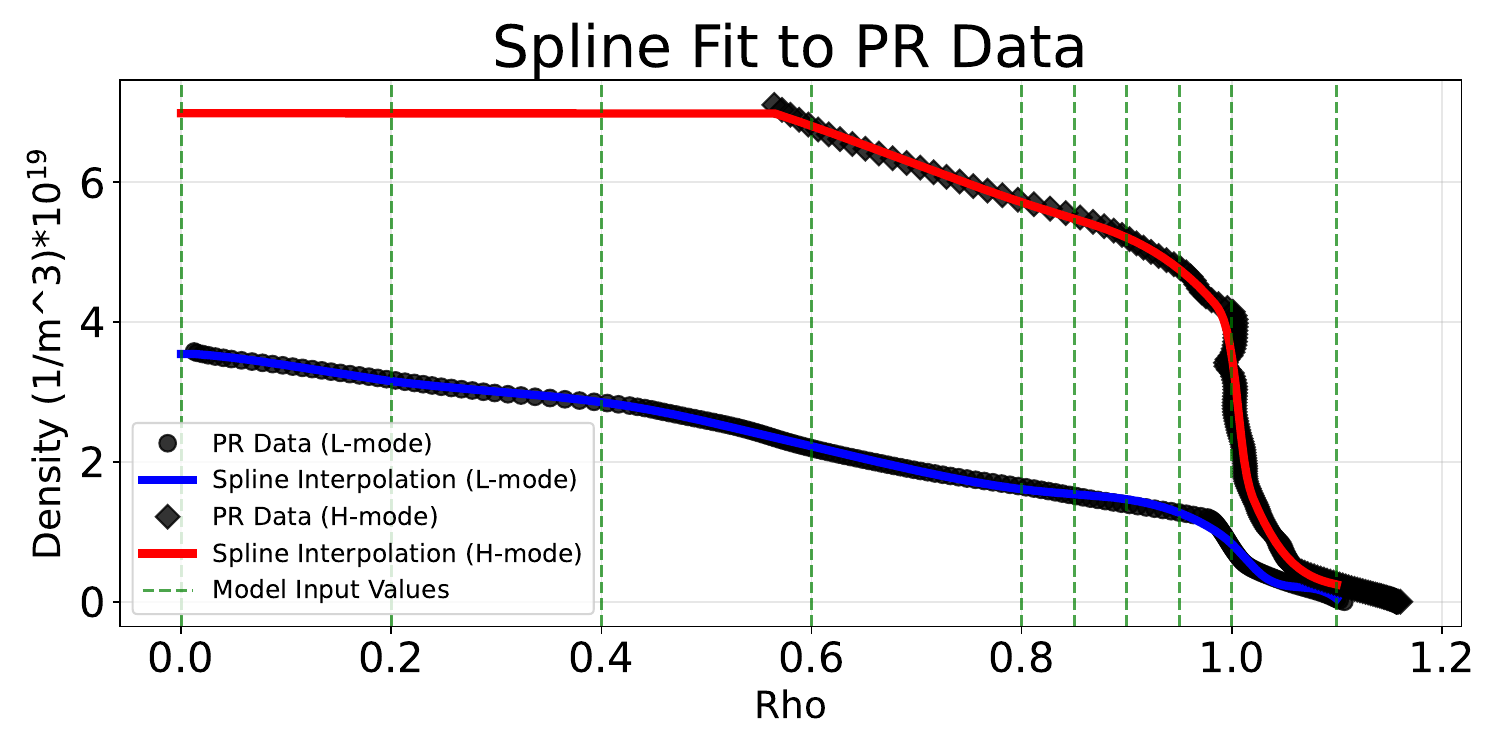}
\caption{This plot is an example of how well the splines fit to the data, how they smooth out the noise, and how they perform the constant extrapolation in the H-mode case where the PR density limit is reached. The dashed vertical lines represent locations where density values will be taken for input to the GBC model.}
\label{SplineFit}
\end{figure}

Difficulty arises in extrapolating the data. As shown in Figure \ref{Histogram}, a variety of situations can occur: PR can cover the complete profile from the core to the edge and beyond, measurements can stop short of the core with $\rho{\text{min}}>0$, and sometimes even stop at the edge $\rho{\text{min}}\simeq 1$. 
An example of partial coverage is shown in Figure \ref{SplineFit}. We have considered complementing the data by inferring ferring the missing core data from the covered region. After experimentation, we have decided against it. Extrapolation to the core is less informative for the H-mode classification task, which primarily deals with the edge region.  If the covered region contains useful information about the core, the classifier should be able to extract it without extrapolation. Instead, we pad the missing data with the deepest known value.  


The inputs encapsulate information about the edge region for pedestal detection, where the density limit is reached (if at all), and leverage the general shape of the density profile. To achieve this, we chose 10 points along the profile, as shown in Figure \ref{SplineFit}, and used the fitted spline model's values at each point as input to the binary classification model. These ten points are located along $\rho$ at $\rho^*$ = [0,0.2,0.4,0.6,0.8,0.85,0.9,0.95,1.0,1.1]: the density of the points is higher at the edge. We have decided against making the number of knots and their locations the subject of an additional optimization: the potential gain from fine-tuning is minimal, but it could make the model brittle. 

\begin{figure}[ht!]
\centering
\includegraphics[width=8.5cm,height=8.5cm,keepaspectratio]{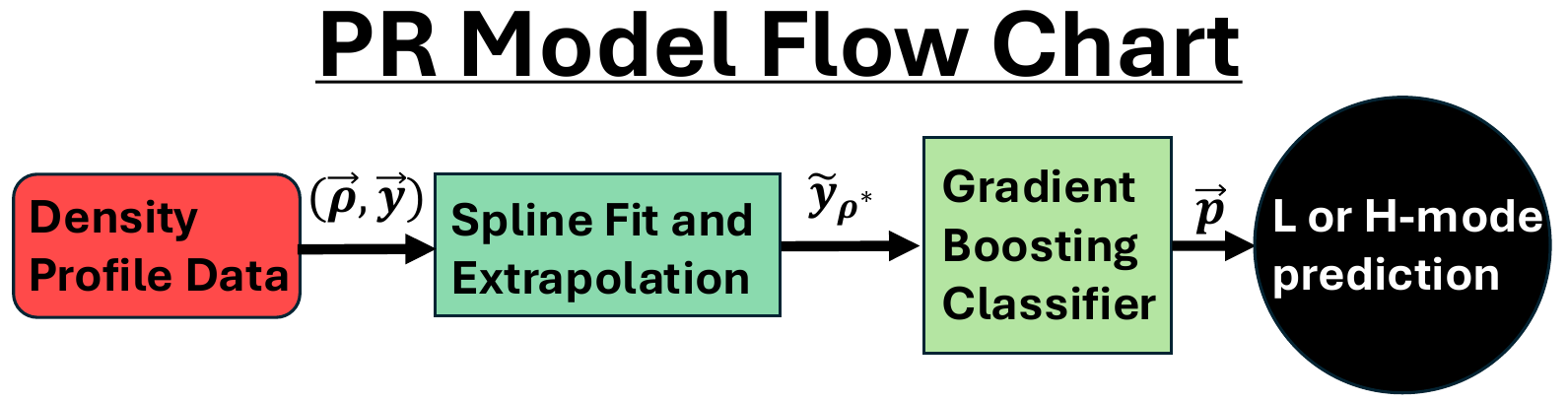}
\caption{This flow chart shows the full PR model, how data from the PR is processed and used for confinement mode classification.}
\label{FlowChart}
\end{figure}

The spline fit outputs at $\rho^*$ are used to train a Gradient Boosted Classifier (GBC) from sklearn, as it is robust, fast, and accurate. The full model, from start to finish, is shown in Figure \ref{FlowChart}.

\begin{table}[!ht]
\begin{center}
\begin{tabular}{||c c||} 
 \hline
 Model Parameter & Value  \\ [0.5ex] 
 \hline\hline
 Total Shots  & 260  \\ 
 \hline
  Train/Test Split  & 80/20  \\ 
 \hline
 Total Data Points  & 8102  \\ 
 \hline
 L:H ratio   & 2762:5340  \\ 
 \hline
  PR Density Limit  & $\sim6.5 \cdot 10^{19}$~m$^{-3}$   \\ 
  \hline
 Polynomial Spline Order & 3 \\ 
 \hline
 Spline Knots & 10 \\ 
 \hline
 Learning Rate & 0.1  \\ 
 \hline
 Max Iterations & 100 \\
 \hline
  Max Leaf Nodes & 31  \\ 
 \hline
  L2 Regularization & 0  \\ 
 \hline
\end{tabular}
\end{center}
\caption{Table of Parameters used for building and curating the data set, fitting the splines, and training the GBC.}
\label{GBC_HyperParam}
\end{table}

\subsection{Model Analysis}
The model's performance is outlined in Table \ref{GBCModelPerformance}, which shows 97\% test accuracy among other statistics. This model has been tested across many reshuffles of different training and test shot arrangements. Snapshot data within each shot are correlated, which reduces the effective number of data points for calibration and testing. For this reason, we have decided to split the test and train data by shots rather than by snapshots. This reduces the risk of overfitting, temporal leakage, and, ultimately, overestimation of model performance. 

\begin{figure}[ht!]
\centering
\includegraphics[width=8.5cm,height=8.5cm,keepaspectratio]{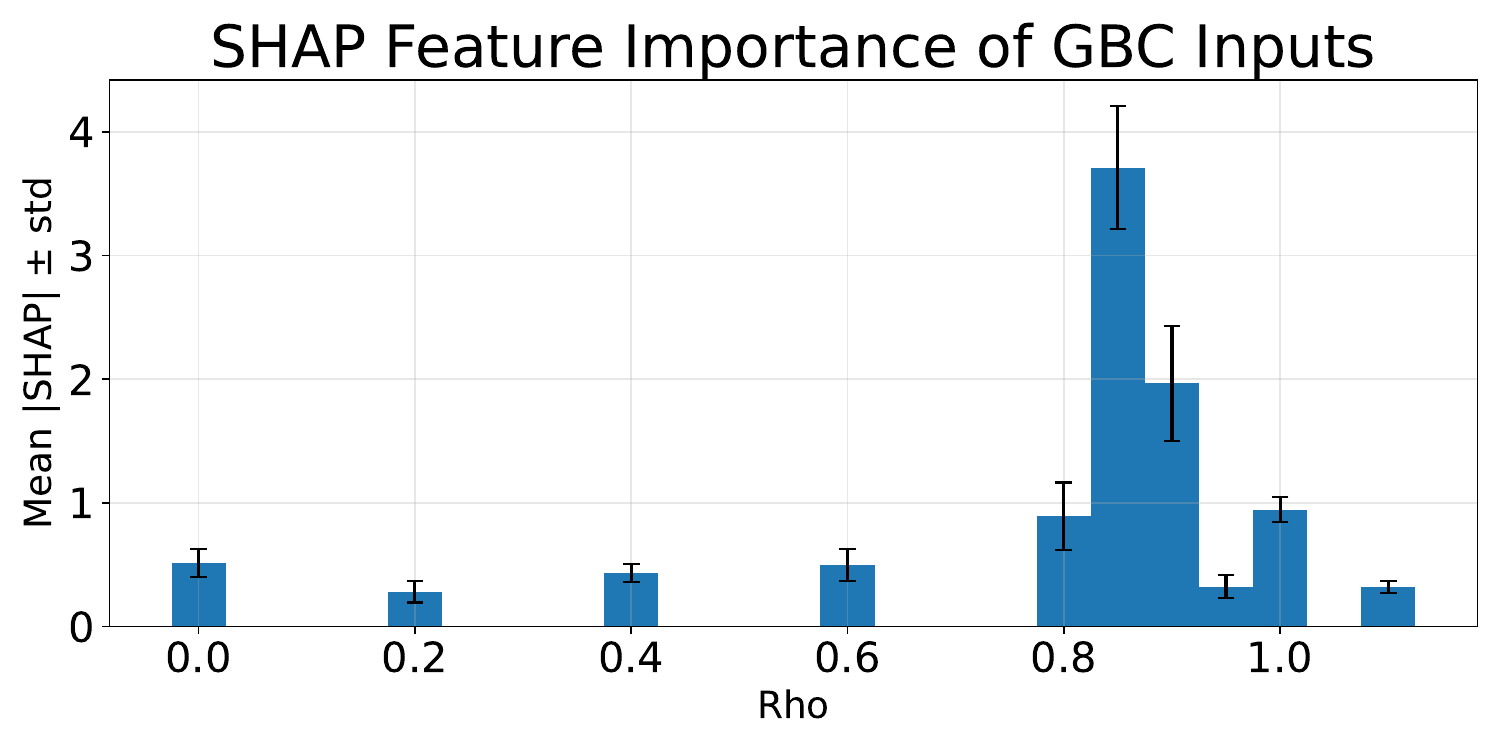}
\caption{This plot depicts the relative importance of each input into the GBC as determined by the location of the density measurement along $\rho$. This result shows the dominant contribution of input data from the edge region, strongly indicating the importance of pedestal detection for identifying H-mode. The physics that informed the construction of this model is identifiable in this plot.}
\label{SHAP}
\end{figure}

To better understand the constructed model and evaluate the significance of each input, we have conducted a Shapley value-based analysis  \cite{winter2002shapley}. Shapley values are computed using a heuristic that roughly estimates each input's relative contribution to the model's overall performance.  The results in Figure \ref{SHAP} describe the relative importance of each model input. It demonstrates that the edge region is most responsible for the model prediction. This result confirms the physics-based intuition that the model's performance is determined mainly by the area around the pedestal. It also supports our decision to increase the density of inputs near the plasma edge, where most of the information about the plasma state comes from.   

\begin{table}[!ht]
\begin{center}
\begin{tabular}{||c c c||} 
 \hline
 PR Model & Average & Standard Deviation \\ [0.5ex] 
 \hline\hline
 Test Accuracy  & 97\% & 1.0\% \\ 
 \hline
 Test Precision  & 98\% & 0.90\% \\ 
 \hline
 Test Recall  & 98\% & 1.5\% \\ 
 \hline
 Test F1  & 98\% & 0.80\% \\ 
 \hline
\end{tabular}
\end{center}
\caption{The statistics of the GBC model using spline-fitted PR data with constant extrapolation as input. The statistics are computed across 100 reshuffles of shots to produce different training and testing sets.}
\label{GBCModelPerformance}
\end{table}

\section{The Ensemble Model}
This paper has so far covered the invention of the PR based H-mode detector. In our previous paper, we discussed an ECE-based H-mode detector. Now we seek to combine the two models into an ensemble to improve identification accuracy and robustness. The combination model must take into account each model's confidence in its prediction, recognize each model's limits, and use this information to drive the choice between the two predictions. It is also helpful for each model to perform rudimentary uncertainty quantification and evaluate whether a particular sample lies within a densely covered region of feature space or is an outlier. 

Both the PR and ECE-based models rely on their respective feature extraction methods. To assign a reliability score to feature vectors, we adopt an approach similar to anomaly detection. We use the k-means clustering algorithm \cite{likas2003global}, in which centers are placed throughout the feature space of the training data set. These centers are located in regions of densely packed training data points and can be used to identify the area of feature space well explored by the model. Test data points that are farther from the centers than the average training data point distance should be weighted less confidently than test data points that are relatively close to these centers. 

\begin{equation}
EM = \frac{C_{PR}(x_{PR})P_{PR}+C_{ECE}(x_{ECE})P_{ECE}}{C_{PR}(x_{PR})+C_{ECE}(x_{ECE})}
\end{equation}
\begin{equation}
C_{PR}(x_{PR}) = \min (\bar{x}_{PR}*c_{PR}/x_{PR},1)
\end{equation}
\begin{equation}
C_{ECE}(x_{ECE}) = \min (\bar{x}_{ECE}*c_{ECE}/x_{ECE},1)
\end{equation}

The ensemble model is represented a weighted average of feature inputs where their confidence weighting, $C_{PR}$ and $C_{ECE}$, are evaluated by taking the minimum value of 1 and a ratio of the average training weight times a coefficient c (found heuristically to work well at 3.5 for the ECE model and 2.5 for the PR model) and the test feature data point. $P_{PR}$ and $P_{ECE}$ are the probabilities outputted by the individual PR and ECE models.

\begin{table}[!ht]
\begin{center}
\begin{tabular}{||c c c||} 
 \hline
 Ensemble Model & Average & Standard Deviation \\ [0.5ex] 
 \hline\hline
 Test Accuracy  & 99.2\% & 0.79\% \\ 
 \hline
 Test Precision  & 98.8\% & 1.3\% \\ 
 \hline
 Test Recall  & 99.6\% & 0.36\% \\ 
 \hline
 Test F1  & 99.2\% & 0.70\% \\ 
 \hline
\end{tabular}
\end{center}
\caption{The statistics of the Ensemble model using the weighted average of the PR and ECE confinement mode classification models. The statistics are taken across 100 reshuffles of shot numbers as the shot numbers, similar to Table \ref{GBCModelPerformance}.}
\label{EnsembleModelPerformance}
\end{table}

Undoubtedly, the ensemble model is a notable improvement over the PR model by itself as indicated in Table \ref{EnsembleModelPerformance}. The approach of combining both models has resulted in a model with significantly higher test accuracy and the ability to use the combined predictions to compensate for times when one individual model is found lacking. 

This raises an interesting and consequential question for FPP-relevant diagnostics: if there exists an unforeseen event, not covered in training (perhaps a large perturbative event, or an unfamiliar plasma regime not covered by the 300 shots), then is it even possible for one feature vector to be anomalous while the other is normal? 

In a series of data explorations, we have found that it is indeed quite common for one test data point to be anomalous while the other diagnostic is normal. Treating as anomalous data points which are 2.5 times farther from a nearest k-means center than the average training distance and normal data points within 1.0 times the average training distance,  we find that when ECE is anomalous, 17\% of the PR dataset is normal. When PR is anomalous, 31\% of the ECE dataset is normal. 

These results confirm that there is indeed value in attaching confidence metrics to the two models, as they can address many of each other's blind spots. 
\section{Robustness to Future Data}
The H-mode identification models developed, the ECE, PR, and Ensemble models, have high test accuracies for shot randomized trials. However, despite these strong results, the question remains about the models' ability to handle future data that may deviate significantly from the training data, which could degrade model performance. The classifiers were designed using feature-extraction methods for pedestal detection to make them robust to changes in the data landscape; this is expected as DIII-D operations evolve rapidly to address different research questions.

Nonetheless, to estimate the model's robustness to future data, we have conducted a sliding-window test, separating the training and test windows into chronologically ordered windows. Of the 300 labeled shots, 200 are used in training, and 60 are used for testing with gap shots before and after the test window to ensure clean separation between the two windows. Due to the limited number of labeled shots, the windows will wrap around, meaning the windows will transition from the last chronological shot to the first. We argue that this is an acceptable approach because with this method, the test window still comprises shots and experiments not related to those in the training window.

\begin{figure}[ht!]
\centering
\includegraphics[width=8.5cm,height=8.5cm,keepaspectratio]{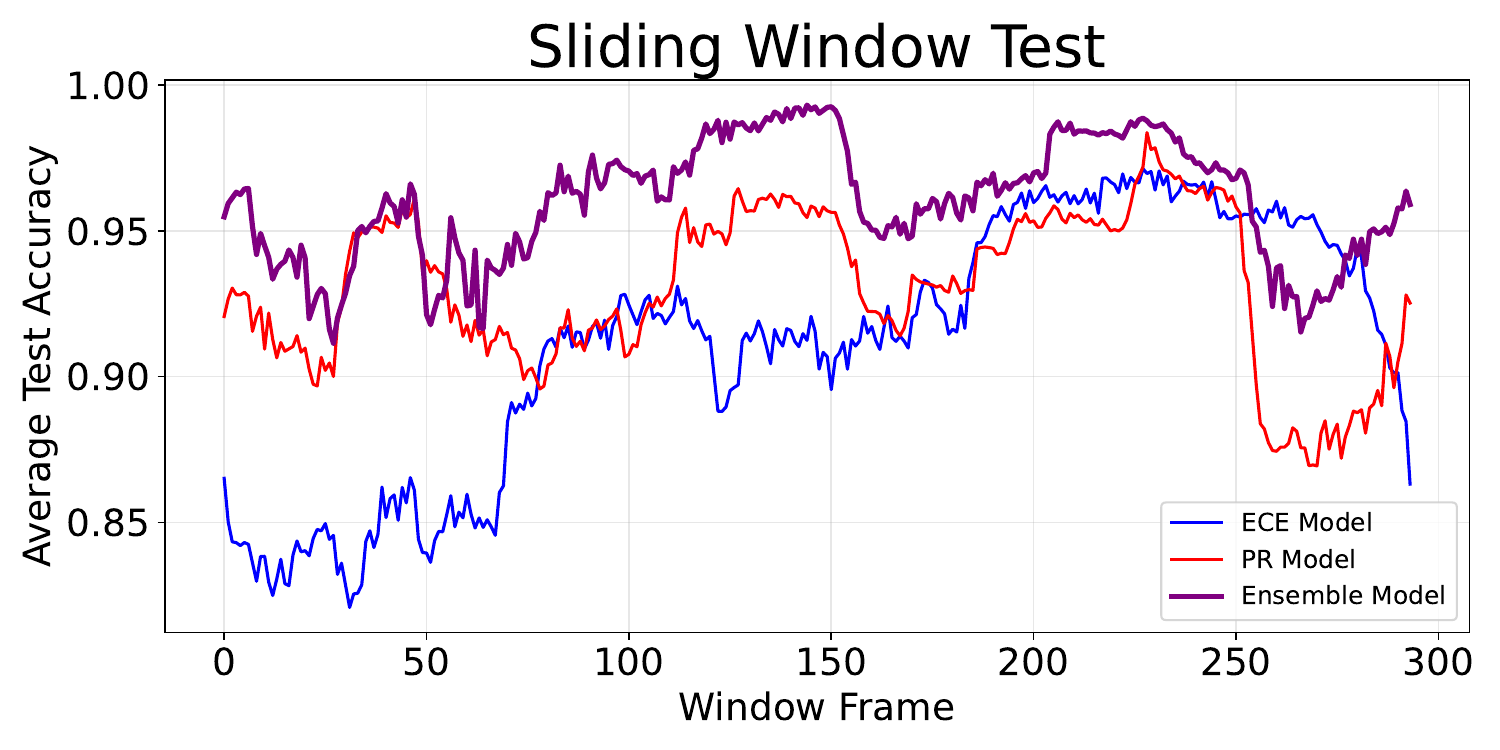}
\caption{This sliding window test breaks up the labeled data set chronologically into 200 training shots, 60 test shots, and gap shots separating the train and test windows. The x-axis shows the sliding window frame, where each new entry represents the training and test windows, adopting a new shot and discarding the last one. The average test accuracy per frame is shown for the ECE, PR, and Ensemble models.}
\label{SlidingWindowPic}
\end{figure}

Figure \ref{SlidingWindowPic} depicts the performance of each model for this test, with the global statistics of the test displayed in Table \ref{SlidingWindowTable}. The results indicate each model's confidence in handling unforeseen data in future experiments. There is a disparity in model performance between the sliding window test and randomized shot tests, which can be attributed to the similarity of data when shots are randomized vs when structured time frames separate them. If more labeled data could be obtained, the training window could encapsulate the full range of diagnostic behavior and the disparity between accuracies would dissipate. Regardless, the limitation of labeled data has not invalidated the existing models, as they still show high performance in predicting the confinement modes relative to other models proposed in the fusion community \cite{gill2024real,matos2021plasma,orozco2022neural}. These tests, however, indicate that for optimal performance, periodic recalibration may be necessary to assimilate new data into the model. 

\begin{table}[!ht]
\begin{center}
\begin{tabular}{||c c c||} 
 \hline
 Model & Average Accuracy & Standard Deviation \\ [0.5ex] 
 \hline\hline
 ECE  & 91.1\% & 4.3\% \\ 
 \hline
 PR  & 93.1\% & 2.6\% \\ 
 \hline
 Ensemble  & 96.0\% & 2.0\% \\ 
 \hline
\end{tabular}
\end{center}
\caption{The results from the sliding window test are used to calculate the global average and standard deviation of each model on future data. The result shows a drop in accuracy in each model compared to randomized tests while still maintaining a high H-mode identification rate, particularly for the ensemble model.}
\label{SlidingWindowTable}
\end{table}

\section{Conclusion and future work}
We have extended the research project goal of developing plasma state classifiers restricted to reactor-relevant diagnostics with a Profile Reflectometer-based confinement mode classifier. We have taken this one step further by proposing an ensemble model of the new PR model and the previous ECE model. The ensemble model achieves accuracies greater than those of either of the two individual models. 

The overall goal of expanding the known capabilities of existing FPP-relevant diagnostics will continue while making the new tools available to the fusion community. This project is of significant importance to the fusion community, as it guides the choice of diagnostics for the limited space FPPs will have for diagnostics. In this paper we have established a benchmark for PR based confinement mode classification accuracy in DIII-D. 

Future work will expand to new plasma classification tasks related to control and fusion performance with new FPP-relevant diagnostics. 
\section{Acknowledgment}
This material is based upon work supported by the U.S. Department of Energy, Office of Science, Office of Fusion Energy Sciences, using the DIII-D National Fusion Facility, a DOE Office of Science user facility, under Awards DE-FC02-04ER54698, DE-FG02-05ER54809, DE-FG02-97ER54415, DE-SC0019352, and Next Step Fusion S.a.r.l. with UCSD staff supported by Next Step Fusion S.a.r.l. The authors would like to thank Terry Rhodes for fruitful discussions.
\section*{Disclaimer}
This report was prepared as an account of work sponsored by an agency of the United States Government. Neither the United States Government nor any agency thereof, nor any of their employees, makes any warranty, express or implied, or assumes any legal liability or responsibility for the accuracy, completeness, or usefulness of any information, apparatus, product, or process disclosed, or represents that its use would not infringe privately owned rights. Reference herein to any specific commercial product, process, or service by trade name, trademark, manufacturer, or otherwise does not necessarily constitute or imply its endorsement, recommendation, or favoring by the United States Government or any agency thereof. The views and opinions of authors expressed herein do not necessarily state or reflect those of the United States Government or any agency thereof.

\bibliographystyle{plain}

\begin{thebibliography}{10}

\bibitem{austin2003}
ME~Austin and J~Lohr.
\newblock Electron cyclotron emission radiometer upgrade on the diii-d tokamak.
\newblock {\em Review of Scientific Instruments}, 74(3):1457--1459, 2003.

\bibitem{clark2025plasma}
Randall Clark, Vacslav Glukhov, Georgy Subbotin, Maxim Nurgaliev, Aleksandr Kachkin, Max Austin, and Dmitri~M Orlov.
\newblock Plasma confinement state classification via fpp relevant microwave diagnostics.
\newblock {\em arXiv preprint arXiv:2510.14078}, 2025.

\bibitem{de1978practical}
Carl De~Boor and Carl De~Boor.
\newblock {\em A practical guide to splines}, volume~27.
\newblock springer New York, 1978.

\bibitem{gill2024real}
Kevin Gill, David Smith, S~Joung, B~Geiger, G~McKee, J~Zimmerman, R~Coffee, A~Jalalvand, and E~Kolemen.
\newblock Real-time confinement regime detection in fusion plasmas with convolutional neural networks and high-bandwidth edge fluctuation measurements.
\newblock {\em Machine Learning: Science and Technology}, 5(3):035012, 2024.

\bibitem{hartfuss2013mwdiag}
HJ~Hartfuss and T~Geist.
\newblock {\em Fusion Plasma Diagnostics with mm-Waves}.
\newblock Wiley-VCH, Weinheim, Germany, 1st edition, 2013.

\bibitem{kim1997development}
KW~Kim, EJ~Doyle, TL~Rhodes, WA~Peebles, CL~Rettig, and NC~Luhmann, Jr.
\newblock Development of a fast solid-state high-resolution density profile reflectometer system on the diii-d tokamak.
\newblock {\em Review of scientific instruments}, 68(1):466--469, 1997.

\bibitem{likas2003global}
Aristidis Likas, Nikos Vlassis, and Jakob~J Verbeek.
\newblock The global k-means clustering algorithm.
\newblock {\em Pattern recognition}, 36(2):451--461, 2003.

\bibitem{matos2021plasma}
Francisco Matos, Vlado Menkovski, Alessandro Pau, Gino Marceca, Frank Jenko, TCV Team, et~al.
\newblock Plasma confinement mode classification using a sequence-to-sequence neural network with attention.
\newblock {\em Nuclear Fusion}, 61(4):046019, 2021.

\bibitem{mukhovatov2003overview}
V~Mukhovatov, M~Shimada, AN~Chudnovskiy, AE~Costley, Y~Gribov, G~Federici, O~Kardaun, AS~Kukushkin, A~Polevoi, VD~Pustovitov, et~al.
\newblock Overview of physics basis for iter.
\newblock {\em Plasma physics and controlled fusion}, 45(12A):A235, 2003.

\bibitem{orozco2022neural}
David Orozco, Brian Sammuli, Jayson Barr, William Wehner, and David Humphreys.
\newblock Neural network-based confinement mode prediction for real-time disruption avoidance.
\newblock {\em IEEE Transactions on Plasma Science}, 50(11):4157--4164, 2022.

\bibitem{rodriguez2022overview}
P~Rodriguez-Fernandez, AJ~Creely, MJ~Greenwald, D~Brunner, SB~Ballinger, CP~Chrobak, DT~Garnier, R~Granetz, ZS~Hartwig, NT~Howard, et~al.
\newblock Overview of the sparc physics basis towards the exploration of burning-plasma regimes in high-field, compact tokamaks.
\newblock {\em Nuclear Fusion}, 62(4):042003, 2022.

\bibitem{siccinio2022development}
M~Siccinio, Jonathan~Peter Graves, R~Kembleton, H~Lux, F~Maviglia, AW~Morris, J~Morris, and H~Zohm.
\newblock Development of the plasma scenario for eu-demo: Status and plans.
\newblock {\em Fusion Engineering and Design}, 176:113047, 2022.

\bibitem{JMLR:v9:vandermaaten08a}
Laurens van~der Maaten and Geoffrey Hinton.
\newblock Visualizing data using t-sne.
\newblock {\em Journal of Machine Learning Research}, 9(86):2579--2605, 2008.

\bibitem{winter2002shapley}
Eyal Winter.
\newblock The shapley value.
\newblock {\em Handbook of game theory with economic applications}, 3:2025--2054, 2002.

\bibitem{zeng2006fast}
L~Zeng, G~Wang, EJ~Doyle, TL~Rhodes, WA~Peebles, and Q~Peng.
\newblock Fast automated analysis of high-resolution reflectometer density profiles on diii-d.
\newblock {\em Nuclear fusion}, 46(9):S677, 2006.

\end{thebibliography}

\end{document}